\documentclass[twocolumn,prl,showpacs]{revtex4}
\usepackage{graphics}

\begin{document}

\title{Back-action cancellation in interferometers by quantum locking}

\author{J.-M. Courty}
\author{A. Heidmann}
\author{M. Pinard}

\affiliation{Laboratoire Kastler Brossel, Case 74, 4 place
Jussieu, F75252 Paris Cedex 05, France} \thanks{Unit\'{e} mixte de
recherche du Centre National de la Recherche Scientifique, de
l'Ecole Normale Sup\'{e}rieure et de l'Universit\'{e} Pierre et
Marie Curie} \homepage{www.spectro.jussieu.fr/Mesure}

\begin{abstract}
We show that back-action noise in interferometric measurements
such as gravitational-waves detectors can be completely suppressed
by a local control of mirrors motion. An optomechanical sensor
with an optimized measurement strategy is used to monitor mirror
displacements. A feedback loop then eliminates radiation-pressure
effects without adding noise. This very efficient technique leads
to an increased sensitivity for the interferometric measurement,
which becomes only limited by phase noise. Back-action
cancellation is furthermore insensitive to losses in the
interferometer.
\end{abstract}

\pacs{42.50.Lc, 04.80.Nn, 03.65.Ta}

\maketitle

Sensitivity in interferometric measurements such as
gravitational-waves detectors \cite{Bradaschia90,Abramovici92} is
ultimately limited by quantum noise of light. Phase fluctuations
introduce noise in the measurement whereas radiation pressure of
light induces unwanted mirrors displacements. Both lead to a
quantum limit and potential applications of squeezed states to
overcome this limit have motivated a large number of works in
quantum optics \cite{Caves81,Jaekel90,Braginsky92,Kimble02}.

It has recently been proposed to enhance the sensitivity in
interferometric measurement by active control of mirrors
displacements \cite{Courty03}. Active control can reduce classical
noise such as thermal noise in cold-damped mechanical systems
\cite{Milatz53,Grassia00,Cohadon99,Pinard00} and may in principle
be used to reduce noise in a quantum regime
\cite{Wiseman95,Mancini00,Courty01}.

The scheme proposed in \cite{Courty03} is based on a local control
of each mirror of the interferometer. An optomechanical sensor
made of a high-finesse cavity monitors the mirror motion. A
feedback loop then locks the mirror at the quantum level, with
respect to the position of the other mirror of the sensor cavity.
The sensor sensitivity is transferred to the interferometric
measurement, resulting in a reduction of back-action noise due to
radiation pressure in the interferometer.

In this paper we show that a similar technique can be used to
completely suppress back-action noise. An optimized measurement
strategy for the optomechanical sensor allows one to freeze the
mirror in an absolute way, leading to a complete elimination of
radiation-pressure noise. We furthermore show that this behavior
is insensitive to the characteristics of the interferometer. In
contrast to injection of squeezed states \cite{Kimble02},
back-action cancellation is still obtained in presence of losses
in the interferometer.

\begin{figure}
{\resizebox{7 cm}{!}{\includegraphics{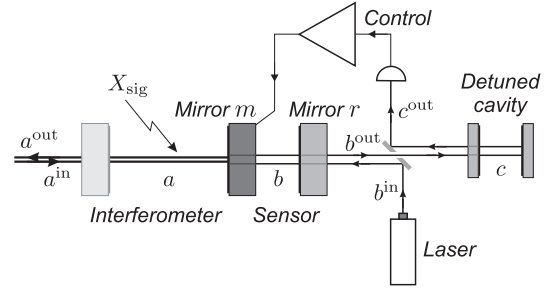}}}
\caption{Scheme of the system studied in the paper. A Fabry-Perot
cavity and a light field $a$ are used to measure a variation
$X_{\rm sig}$ of the cavity length. Mirror $m$ is actively
controlled by a feedback loop via the position measurement
delivered by an optomechanical sensor (cavity made of mirrors $m$
and $r$ with a light field $b$, followed by a detuned cavity).}
\label{Fig_Setup}
\end{figure}

The basic setup is shown in Fig. \ref{Fig_Setup}. We focus on the
active control of one mirror of the interferometer (mirror $m$ in
Fig. \ref{Fig_Setup}). The interferometric measurement is
schematized as the measurement of a length variation $X_{{\rm
sig}}$ of a Fabry-Perot cavity which can be considered as one of
the two arms of a gravitational-wave detector (left part of Fig.
\ref{Fig_Setup}). Motion of mirror $m$ relatively to the position
of a reference mirror $r$ is measured by an optomechanical sensor
made of both mirrors. The intensity of the field reflected by this
high-finesse cavity is measured after being phase-shifted by a
detuned cavity. The result of the measurement is fed back to the
mirror in order to control its displacements.

Fields $a$ and $b$ in both the interferometer and sensor are
described by quantum annihilation operators $a\left[ \Omega
\right] $, $b\left[ \Omega \right] $ at frequency $\Omega $,
whereas mean fields are characterized by complex amplitudes
$\alpha $, $\beta $ normalized in such a way that $\left| \alpha
\right| ^{2}$ and $\left| \beta \right| ^{2}$ correspond to photon
fluxes \cite{Reynaud92}. We define in a usual way the quadrature
$a_{\theta }$ of field $a$ as,
\begin{equation}
a_{\theta }\left[ \Omega \right] =e^{-i\theta }a\left[ \Omega
\right] +e^{i\theta }a^{\dagger }\left[ \Omega \right] .
\end{equation}
When the mean field $\alpha $ is real, intensity and phase
quadratures respectively correspond to the quadrature $a_{0}$
aligned with the mean field and to the orthogonal quadrature
$a_{\frac{\pi }{2}}$.

For a lossless and resonant single-ended cavity, the incident,
intracavity, and reflected mean fields can be taken real. Assuming
the frequency $\Omega $ of interest smaller than the cavity
bandwidth, the input-output relations for the interferometer
cavity are given by \cite{Courty01},
\begin{eqnarray}
\gamma _{{\rm a}}a &=&\sqrt{2\gamma _{{\rm a}}}a^{{\rm
in}}+2ik_{0}\alpha \left( X_{{\rm m}}+X_{{\rm sig} }\right)
,\label{Equ_a}\\
 a^{{\rm out}} &=&-a^{{\rm in}}+\sqrt{2\gamma
_{{\rm a}}}a, \label{Equ_aout}
\end{eqnarray}
where $\gamma _{{\rm a}}$ is the damping rate of the cavity,
$k_{0}$ the field wavevector and $X_{{\rm m}}$ the displacement of
mirror $m$. The intensity quadrature is left unchanged by the
cavity ($a_{0}^{{\rm out}}=a_{0}^{{\rm in }}$) whereas the
input-output phase-shift is proportional to the cavity length
variation,
\begin{equation}
a_{\frac{\pi }{2}}^{{\rm out}}=a_{\frac{\pi }{2}}^{{\rm in}}+2\xi
_{{\rm a}}\left( X_{{\rm m}}+X_{{\rm sig} }\right) .
\end{equation}
The optomechanical coupling parameter $\xi _{{\rm a}}$ is related
to the intracavity mean field amplitude $\alpha $ and to the
cavity finesse ${\cal F}_{ {\rm a}}=\pi /\gamma _{{\rm a}}$,
\begin{equation}
\xi _{{\rm a}}=2k_{0}\alpha \sqrt{2{\cal F}_{{\rm a}}/\pi }.
\label{Equ_xia}
\end{equation}

Measurement of the phase of the reflected field provides an
estimator $\hat{X}_{{\rm sig}}$ of the signal, obtained by a
normalization of the output phase $a_{\frac{\pi }{2}}^{{\rm out}}$
as a displacement. $\hat{X}_{{\rm sig}}$ is the sum of the signal
$X_{{\rm sig}}$ and extra noise terms,
\begin{equation}
\hat{X}_{{\rm sig}}=\frac{1}{2\xi _{{\rm a}}}a_{\frac{\pi
}{2}}^{{\rm out}}=X_{{\rm sig}}+\frac{1}{2\xi _{{\rm
a}}}a_{\frac{\pi }{2}}^{{\rm in}}+X_{{\rm m}}. \label{Equ_Xsig}
\end{equation}
The first noise term is related to the incident phase-noise
$a_{\frac{\pi }{2}}^{{\rm in}}$ and corresponds to the measurement
noise. The second term is the displacement $X_{{\rm m}}$ of mirror
$m$. For a quantum-limited interferometer without control, it
corresponds to the back-action noise due to radiation pressure of
intracavity field $a$. It is deduced from the evolution of the
velocity $V_{{\rm m}}=-i\Omega X_{{\rm m}}$ which can be expressed
from eq. (\ref{Equ_a}) in terms of the incident intensity
quadrature $ a_{0}^{{\rm in}}$,
\begin{equation}
Z_{{\rm m}}V_{{\rm m}}=2\hbar k_{0}\alpha a_{0}=\hbar \xi _{{\rm
a}}a_{0}^{{\rm in}},
\end{equation}
where $Z_{{\rm m}}$ is the mechanical impedance of mirror $m$. At
frequency relevant for gravitational-wave interferometers, a
suspended mirror can be considered as a free mass with an
impedance related to the mirror mass $M_{{\rm m}}$,
\begin{equation}
Z_{{\rm m}}\simeq -i\Omega M_{{\rm m}}. \label{Equ_Zm}
\end{equation}

\begin{figure}
{\resizebox{5.5 cm}{!}{\includegraphics{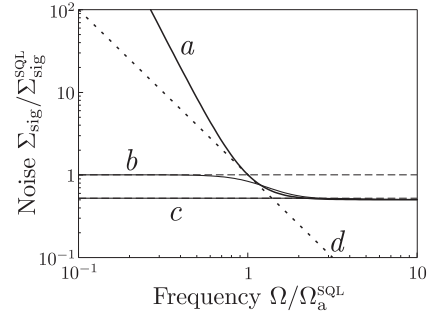}}}
\caption{Equivalent
input noise $\Sigma _{{\rm sig}}$ in the interferometric measurement as a
function of frequency $\Omega $. Curve {\it a}: free interferometer. Curves
{\it b} and {\it c}: control with an optomechanical coupling $\xi _{{\rm b}}$
equal to $\xi _{{\rm a}}$ and $5\xi _{{\rm a}}$, respectively; dashed curves
correspond to an infinite gain and solid lines to the optimum gain. Curve
{\it d}: standard quantum limit. Frequency is normalized to the \textsc{sql}
frequency $\Omega _{{\rm a}}^{\text{{\sc sql}}}$ and noise to $\Sigma _{{\rm
sig}}^{\text{{\sc sql}}}=1/2\xi _{{\rm a}}^{2}$.} \label{Fig_SigOpt}
\end{figure}

Both noises in (\ref{Equ_Xsig}) are uncorrelated for an incident
coherent state, the noise spectra for any quadrature $a_{\theta
}^{{\rm in}}$ being given by \cite{Reynaud92},
\begin{equation}
\sigma _{a_{\theta }a_{\theta }}^{{\rm in}}=1,\quad \sigma
_{a_{\theta }a_{\theta +\frac{\pi }{2}}}^{{\rm in}}=0,
\label{Equ_Saa}
\end{equation}
where the spectrum $\sigma _{a_{\theta }a_{\theta ^{\prime
}}}^{{\rm in}}$ is the quantum average of the symmetrized product
of quadratures,
\begin{equation}
\left\langle a_{\theta }^{{\rm in}}\left[ \Omega \right] \cdot
a_{\theta ^{\prime }}^{{\rm in}} \left[ \Omega ^{\prime }\right]
\right\rangle =2\pi \delta \left( \Omega +\Omega ^{\prime }\right)
\sigma _{a_{\theta }a_{\theta ^{\prime }}}^{{\rm in}}\left[ \Omega
\right] .
\end{equation}
The sensitivity of the interferometer is described by the
equivalent input noise $\Sigma _{{\rm sig}}$ equal to the spectrum
of noises in the estimator $\hat{X}_{{\rm sig}}$. One gets for a
free interferometer,
\begin{equation}
\Sigma _{{\rm sig}}^{{\rm free}}=\frac{1}{4\xi _{{\rm a}}^{2}} \left[
1+\left( \Omega _{{\rm a}}^{\text{{\sc sql}}}/\Omega \right) ^{4}\right] ,
\label{Equ_Ssigfree}
\end{equation}
where $\Omega _{{\rm a}}^{\text{{\sc sql}}}$ is the frequency
where contributions of both noises are equal,
\begin{equation}
\Omega _{{\rm a}}^{\text{{\sc sql}}}=\sqrt {\frac{2\hbar \xi
_{{\rm a}}^{2}}{M_{{\rm m}}}}. \label{Equ_Oasql}
\end{equation}
As shown in curve {\it a} of Fig. \ref{Fig_SigOpt}, phase noise is
dominant at high frequency with a flat frequency dependence,
whereas radiation pressure is dominant at low frequency with a
$1/\Omega ^{4}$ dependence. This behavior leads to the so-called
standard quantum limit for a free interferometer with incident
coherent light \cite{Caves81,Jaekel90,Braginsky92}. It corresponds
to the minimum noise level reachable at a given frequency by
varying the optomechanical coupling $\xi _{{\rm a}}$ (curve {\it
d}).

The sensor cavity measures the motion of mirror $m$. Since it is a
resonant single-ended cavity, field $b$ obeys equations similar to
(\ref{Equ_a}) and (\ref{Equ_aout}) except for the cavity length
variation now equal to $X_{{\rm r}}-X_{{\rm m} }$ ($X_{{\rm r}}$
is the displacement of the reference mirror $r$). For this
measurement we take advantage of the squeezing of light due to the
self phase-modulation induced by radiation pressure
\cite{Kimble02}. Instead of measuring the phase quadrature
$b_{\frac{\pi }{2}}^{{\rm out}}$, we detect a properly chosen
quadrature $b_{\theta }^{{\rm out}}$ of the reflected field.
Equations for this quadrature and for mirror $r$ are,
\begin{eqnarray}
b_{\theta }^{{\rm out}} &=&b_{\theta }^{{\rm in}}+2\xi _{{\rm
b}}\sin \theta \left( X_{{\rm r}}-X_{{\rm m} }\right) ,\\
Z_{{\rm r}}V_{{\rm r}} &=&\hbar \xi _{{\rm b}}b_{0}^{{\rm in}},
\end{eqnarray}
where the optomechanical coupling $\xi _{{\rm b}}$ for cavity $b$
is defined in the same way as $\xi _{{\rm a}}$ [eq.
(\ref{Equ_xia})]. Note that values of $\xi _{{\rm b}}$ as large as
$\xi _{{\rm a}}$ are experimentally accessible with moderate
incident power by using a high-finesse cavity \cite{Courty03}. In
the following, mirror $r$ is for simplicity assumed to be
identical to mirror $m$ so that $Z_{{\rm r}}=Z_{{\rm m}}$.

The sensor provides an estimator $\hat{X}_{{\rm m}}$ of the
displacement of mirror $m$. The sensitivity for this measurement
is limited by the phase noise of beam $b$ and by the motion of
mirror $r$ due to radiation pressure,
\begin{eqnarray}
\hat{X}_{{\rm m}} &=&-\frac{1}{2\xi _{{\rm b}}\sin \theta
}b_{\theta }^{{\rm out}}  \nonumber \\
&=&X_{{\rm m}}-\frac{1}{2\xi _{{\rm b}}}b_{\frac{\pi }{2}}^{{\rm
in}}-\left( \frac{\cot \theta }{2\xi _{{\rm b}}}+\frac{i\hbar \xi
_{{\rm b}}}{\Omega Z_{{\rm r}}}\right) b_{0}^{{\rm in}}.
\label{Equ_Xmest}
\end{eqnarray}
Last term vanishes for a frequency-dependent value $\theta ^{{\rm opt}}$ of
quadrature angle $\theta $ defined by,
\begin{equation}
\cot \theta ^{{\rm opt}}=\left( \Omega _{{\rm b}}^{\text{{\sc
sql}}}/\Omega \right) ^{2}, \label{Equ_ThetaOpt}
\end{equation}
where $\Omega _{{\rm b}}^{\text{{\sc sql}}}$ is the \textsc{sql}
frequency for cavity $b$ defined in a similar way as $\Omega
_{{\rm a}}^{\text{{\sc sql}}}$ [eq. (\ref{Equ_Oasql})]. The
equivalent input noise for the sensor measurement is then only
related to incident phase-noise and no longer depends on
radiation-pressure effects on reference mirror $r$. Quadrature
$\theta ^{{\rm opt}}$ actually corresponds to the best strategy
for the signal-to-noise ratio of the measurement \cite{Kimble02}.
At high frequency, radiation-pressure effects are negligible and
the optimal quadrature is the phase $b_{\frac{\pi }{2}}^{{\rm
out}}$ of the reflected field. As frequency decreases, radiation
pressure becomes dominant and the optimal quadrature tends toward
the intensity quadrature.

The result of the sensor measurement is fed back to mirror $m$ via
a force proportional to the estimator $\hat{X}_{{\rm m}}$. The
mirror motion in presence of feedback depends on radiation
pressure of both cavities and on the feedback force,
\begin{equation}
Z_{{\rm m}}V_{{\rm m}}=\hbar \xi _{{\rm a}}a_{0}^{{\rm in}}-\hbar
\xi _{{\rm b}}b_{0}^{{\rm in}}+i\Omega Z_{{\rm fb}}\hat{X}_{{\rm
m}}, \label{Equ_Vm}
\end{equation}
where $Z_{{\rm fb}}$ is the transfer function of the feedback
loop. For the optimum detection strategy [eq.
(\ref{Equ_ThetaOpt})], the resulting motion of mirror $m$ is,
\begin{equation}
\left( Z_{{\rm m}}+Z_{{\rm fb}}\right) V_{{\rm m}}=\hbar \xi
_{{\rm a} }a_{0}^{{\rm in}}-\hbar \xi _{{\rm b}}b_{0}^{{\rm
in}}-\frac{i\Omega }{2\xi _{{\rm b}}}Z_{{\rm fb}}b_{\frac{\pi
}{2}}^{{\rm in}}. \label{Equ_Vmfb}
\end{equation}
The main effect of control is to change the response of mirror $m$ to
radiation-pressure by adding a feedback-induced impedance $Z_{{\rm fb}}$ to
the free mechanical impedance $Z_{{\rm m}}$. For a large feedback gain, the
effective impedance is increased, then reducing mirror displacements. The
control also contaminates mirror displacements by the noise in the sensor
measurement [last term in eq. (\ref{Equ_Vmfb})]. Since this noise is only
related to incident phase-noise of cavity $b$, the control can freeze the
motion of mirror $m$ in an absolute way, down to the limit associated to this
noise. In contrast to a non-optimized measurement \cite{Courty03}, the mirror
locking is insensitive to the motion of reference mirror $r$ induced by
radiation-pressure fluctuations.

One gets from eq. (\ref{Equ_Xsig}) the estimator $\hat{X}_{{\rm sig}}$ for
the interferometer in presence of feedback,
\begin{eqnarray}
\hat{X}_{{\rm sig}} &=&X_{{\rm sig}}+\frac{1}{2\xi _{{\rm
a}}}a_{\frac{\pi }{2}}^{ {\rm in}}+\frac{1}{2\xi _{{\rm
b}}}\frac{Z_{{\rm fb}}}{Z_{{\rm m}}+Z_{{\rm fb}}}b_{\frac{\pi
}{2}}^{{\rm in}} \nonumber \\
&&+\frac{i\hbar }{\Omega \left( Z_{{\rm m}}+Z_{{\rm fb}}\right)
}\left( \xi _{{\rm a}}a_{0}^{{\rm in}}-\xi _{{\rm b}}b_{0}^{{\rm
in}}\right) .
\end{eqnarray}
Since all noises are uncorrelated the equivalent input noise is
given by,
\begin{equation}
\Sigma _{{\rm sig}}=\frac{1}{4\xi _{{\rm a}}^{2}}+\frac{1}{4\xi
_{{\rm b}}^{2}}\left| \frac{Z_{{\rm fb}}}{Z_{{\rm m}}+Z_{{\rm
fb}}}\right| ^{2}+\frac{\hbar ^{2}\left( \xi _{{\rm a}}^{2}+\xi
_{{\rm b}}^{2}\right) }{\Omega ^{2}\left| Z_{{\rm m}}+Z_{{\rm
fb}}\right| ^{2}}. \label{Equ_Ssig}
\end{equation}
For an infinite feedback gain, radiation-pressure noise is completely
suppressed [last term in eq. (\ref{Equ_Ssig})] and the equivalent input noise
reduces to the sum $1/4\xi _{{\rm a}}^{2}+1/4\xi _{{\rm b}}^{2}$ of phase
noises of both cavities. The sensitivity no longer depends on frequency and
tends to the phase noise of the interferometer alone as $\xi _{{\rm b}}$
increases (dashed curves in Fig. \ref{Fig_SigOpt}). The best sensitivity is
obtained by optimizing the gain in eq. (\ref{Equ_Ssig}),
\begin{eqnarray}
Z_{{\rm fb}}^{{\rm opt}} &=&Z_{{\rm m}}\left( \Omega _{{\rm fb}}/
\Omega \right) ^{4}, \label{Equ_Zfbopt} \\
\Sigma _{{\rm sig}}^{{\rm opt}} &=&\frac{1}{4\xi _{{\rm
a}}^{2}}+\frac{1}{4\xi _{{\rm b}}^{2}\left[ 1+\left( \Omega
/\Omega _{{\rm fb}}\right) ^{4}\right]}, \label{Equ_Ssigopt}
\end{eqnarray}
where frequency $\Omega _{{\rm fb}}$ is defined as,
\begin{equation}
\Omega _{{\rm fb}}^{2}=\Omega _{{\rm b}}^{\text{{\sc
sql}}}\sqrt{\left( \Omega _{{\rm a}}^{\text{{\sc sql}}}\right)
^{2}+\left( \Omega _{{\rm b}}^{\text{{\sc sql}}}\right) ^{2}}.
\end{equation}
For frequencies smaller than the cutoff frequency $\Omega _{{\rm fb}}$, the
resulting noise is similar to the one obtained for an infinite gain, whereas
it reduces to the phase noise of the interferometer alone at high frequency.
As shown in Fig. \ref{Fig_SigOpt}, one gets a clear increase of sensitivity
at low frequency where back-action noise is completely suppressed and
replaced by the phase noise of field $b$, without any loss at high frequency.

The optimized detection strategy [eq. (\ref{Equ_ThetaOpt})] is
achieved by sending the field $b^{{\rm out}}$ in a detuned cavity
and by detecting the reflected intensity (see Fig.
\ref{Fig_Setup}). For a rigid and detuned cavity with a bandwidth
comparable to frequency of interest, expression (\ref{Equ_a}) of
the intracavity field is modified as,
\begin{equation}
\left( \gamma +i\Delta -i\Omega \tau \right) c=\sqrt{2\gamma
}c^{{\rm in}}=\sqrt{2\gamma }b^{{\rm out}},
\end{equation}
where $\Delta $ is the detuning and $\tau $ the round trip time of the
cavity. According to the input-output relation (\ref{Equ_aout}) written for
field $c$, the cavity simply induces a frequency-dependent rotation in phase
space of quadratures, the output quadrature $c_{\theta }^{{\rm out}}\left[
\Omega \right] $ being equivalent to the input quadrature $b_{\theta -\phi
_{\Omega }}^{{\rm out}}\left[ \Omega \right] $ with a rotation angle $\phi
_{\Omega }$ given by,
\begin{equation}
\cot \phi _{\Omega }=\frac{\gamma ^{2}-\Delta ^{2}+\Omega ^{2}\tau
^{2}}{-2\gamma \Delta }.
\end{equation}
Since the mean reflected field is rotated by an angle $\phi _{0}$, the
measured intensity quadrature $c_{\phi _{0}}^{{\rm out}}\left[ \Omega \right]
$ corresponds to the quadrature $b_{\phi _{0}-\phi _{\Omega }}^{{\rm
out}}\left[ \Omega \right] $. One gets the correct angle $\phi _{0}-\phi
_{\Omega }=\theta ^{{\rm opt}}$ by choosing the cavity bandwidth $\gamma
/\tau $ and the detuning $\Delta$ as,
\begin{equation}
\gamma /\tau =-\Delta /\tau =\Omega _{{\rm b}}^{\text{{\sc
sql}}}/\sqrt{2}.
\end{equation}
Taking into account the global phase-shift experienced by the field in the
detuned cavity, the input signal $\hat{X}_{{\rm m}}$ of the feedback loop
[eq. (\ref{Equ_Xmest})] is given by,
\begin{equation}
\hat{X}_{{\rm m}}\left[ \Omega \right] =-\frac{1}{2\xi _{{\rm b}}}
\frac{\left( \Omega _{{\rm b}}^{\text{{\sc sql}}}\right)
^{2}-\Omega ^{2}-\sqrt{2}i\Omega _{{\rm b}}^{\text{{\sc
sql}}}\Omega }{\Omega ^{2}}c_{\phi _{0}}^{{\rm out}}\left[ \Omega
\right] .
\end{equation}
It corresponds to a causal filtering of the intensity fluctuations reflected
by the detuned cavity.

We finally analyze the effects of optical losses on control performances.
Losses in the interferometer can be accounted for by an additional damping
coefficient $\gamma _{{\rm v}}$ for the cavity and by a coupling to a vacuum
field $v^{{\rm in}}$ \cite{Reynaud92}. Eq. (\ref{Equ_a}) is modified to,
\begin{eqnarray}
\left( \gamma _{{\rm a}}+\gamma _{{\rm v}}\right) a
&=&\sqrt{2\gamma _{{\rm a}}}a^{{\rm in}}+\sqrt{2\gamma _{{\rm
v}}}v^{{\rm in}} \nonumber \\
&&+2ik_{0}\alpha \left( X_{{\rm m}}+X_{{\rm sig} }\right) .
\label{Equ_aloss}
\end{eqnarray}
Proportion of loss is defined by the coefficient $\eta _{{\rm a}}=\gamma
_{{\rm v}}/\left( \gamma _{{\rm a}}+\gamma _{{\rm v}}\right) $. Losses in the
sensor cavity are described by a similar equation with a coefficient $\eta
_{{\rm b}}$. For the same detection strategy as previously [eq.
(\ref{Equ_ThetaOpt})], one can derive the optimum gain and the interferometer
sensitivity,
\begin{eqnarray}
Z_{{\rm fb}}^{{\rm opt}} &=&Z_{{\rm m}}\left( 1-\eta _{{\rm b}}\right)
\frac{\Omega _{{\rm fb}}^{4}}{\Omega ^{4}+\eta _{{\rm b}}\left( \Omega
_{{\rm b}}^{\text{{\sc sql}}}\right) ^{4}}, \label{Equ_Zfbloss} \\
\Sigma _{{\rm sig}}^{{\rm opt}} &=&\frac{1}{4\xi _{{\rm
a}}^{2}\left( 1-\eta _{{\rm a}}\right) } \nonumber \\
&&+\frac{1}{4\xi _{{\rm b}}^{2}}\left[ \frac{1-\eta _{{\rm
b}}}{1+\eta _{{\rm b}}\left( \Omega _{{\rm b}}^{\text{{\sc
sql}}}/\Omega \right) ^{4}}+\left( \Omega /\Omega _{{\rm
fb}}\right) ^{4}\right] ^{-1} \label{Equ_Ssigloss}
\end{eqnarray}
which has to be compared to the sensitivity of the free interferometer in
presence of loss, still given by eq. (\ref{Equ_Ssigfree}) with the
phase-noise term $1$ in the bracket replaced by $1/\left( 1-\eta _{{\rm
a}}\right) $. Losses in the interferometer ($\eta _{{\rm a}}>0$, $\eta _{{\rm
b}}=0$) do not affect the control. One still has a complete suppression of
back-action noise since only the phase noise due to field $a$ is modified, in
the same proportion for the free and controlled interferometers.

\begin{figure}
{\resizebox{5.5 cm}{!}{\includegraphics{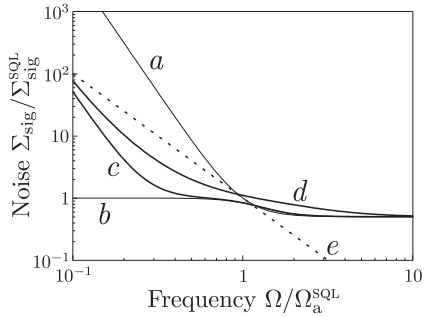}}}
\caption{Equivalent input
noise $\Sigma _{{\rm sig}}$ in the interferometric measurement as
a function of frequency $\Omega $. Curve {\it a}: free
interferometer. Curve {\it b}: control without loss. Curves {\it
c} and {\it d}: control with 1\% loss in the sensor cavity ($\eta
_{{\rm b}}=0.01$), for the optimum feedback gain and for a single
integrator transfer function, respectively. Curve {\it e}:
standard quantum limit. Optomechanical couplings are equal ($\xi
_{{\rm b}}=\xi _{{\rm a}}$).} \label{Fig_SigLoss}
\end{figure}

Cancellation of back-action noise is of course more sensitive to
imperfections in the sensor measurement ($\eta _{{\rm b}}>0$).
Losses induce an additional cutoff frequency $\sqrt[4]{\eta _{{\rm
b}}}\Omega _{{\rm b}}^{\text{{\sc sql}}}$ both for the optimum
gain and the sensitivity [compare eqs. (\ref{Equ_Zfbloss}),
(\ref{Equ_Ssigloss}) to (\ref{Equ_Zfbopt}), (\ref{Equ_Ssigopt})].
At lower frequency the sensitivity can be approximated to,
\begin{equation}
\Sigma _{{\rm sig}}^{{\rm opt}}\simeq \eta _{{\rm b}}\frac{1}{4\xi
_{{\rm b}}^{2}}\left( \Omega _{{\rm b}}^{\text{{\sc sql}}}/\Omega
\right) ^{4}.
\end{equation}
Cancellation is no longer perfect at low frequency. The sensitivity is
contaminated by radiation-pressure effects in the sensor cavity, in
proportion to the loss $\eta _{{\rm b}}$. As shown in Fig. \ref{Fig_SigLoss},
however, one still has a reduction of back-action noise by a factor $100$ for
a $1$\% loss. Curve {\it d} finally shows the sensitivity obtained with a
very simple implementation of the feedback loop which consists in a single
integrator $Z_{{\rm fb}}/Z_{{\rm m}}=2\Omega _{{\rm a}}^{\text{{\sc
sql}}}/\left( -i\Omega \right) $.

To conclude, a local control of mirrors with an optimized
measurement strategy allows one to completely suppress back-action
noise in interferometric measurement. This very efficient
technique presents the advantage to decouple the constraints
needed to manipulate quantum noise from the interferometer
characteristics. As usual in quantum optics, losses must be
avoided in the optomechanical sensor. Quantum locking is however
insensitive to losses in the interferometer and does not imply any
additional constraint to the interferometer design.

\end{document}